\documentclass[12pt]{article}

\usepackage{graphicx}

\newcommand{\AmS}{{\protect\the\textfont2
  A\kern-.1667em\lower.5ex\hbox{M}\kern-.125emS}}

\begin{document}

\resizebox{0.2\textwidth}{!}{\includegraphics{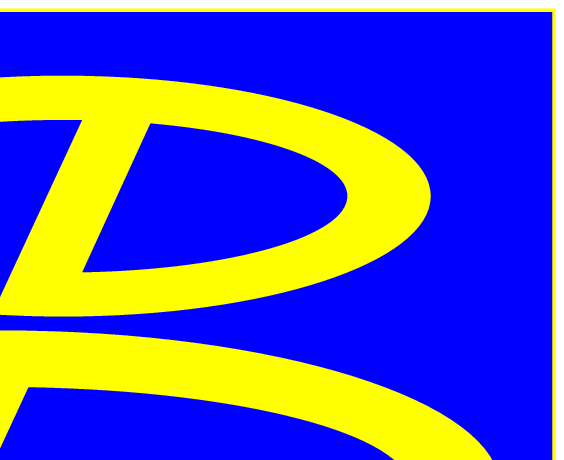}}

\vskip -3cm
\noindent
\hspace*{3.5in}BELLE Preprint 2002-3 \\
\hspace*{3.5in}KEK Preprint 2001-168 \\
\begin{center}
\vskip 2cm
{\Large  \bf
Results of time evolution analyses of $B$-decays at Belle }

\vskip 1cm

{\large Y.~Sakai \\
 \smallskip
  High Energy Accelerator Research Organization (KEK) \\
        1-1, Oho, Tsukuba-shi, Ibaraki-ken, Japan 305-0801 }

\vskip 1cm
\end{center}

\begin{abstract}
\vspace{1pc}
 We report results of proper-time evolution analyses of 
$B$ decays at Belle, based on a 29.1 fb$^{-1}$ data sample recorded
at the $\Upsilon (4S)$ resonance.
These results include
measurements of $\sin2\phi_1$, the lifetime of $B$-mesons,
and the mass-difference between two mass eigenstates of 
the $B^0$-$\bar B^0$ system ($\Delta m_d$).
$\Delta m_d$ is measured using three different methods: fully reconstructed
hadronic modes, a fully reconstructed semi-leptonic mode, and a partially
reconstructed $D^{*\pm}\pi^\mp$ mode.
\end{abstract}


\vspace{5cm}
\begin{center}
 Contributed to the Proceedings of the \\
 Fifth KEK Topical Conference, 20-22, Nov. 2001, Tsukuba, Japan
\end{center}

\newpage

\section{Introduction}

The main goal of a $B$-factory is to 
first observe and establish $CP$ violation in $B$-decays 
and then to precisely measure the $CP$ violation angles of the
Unitarity Triangle, as well as the lengths of its sides.
The aims is to either confirm the
Kobayashi-Maskawa model\cite{KM} for the mechanism of $CP$ violation or 
find evidence of physics beyond the Standard Model. 
Analysis of the proper-time evolution of $B$ decays is 
essential for measuring mixing-induced $CP$ violation,
and is one of the unique
features of an asymmetric $e^+e^-$ collider at the $\Upsilon (4S)$.
By combining a large number of accumulated $B$ events with a
clean environment to study $B$-decays,
the asymmetric-collider experiments can now provide the most precise
measurements that involve proper-decay time analyses, such as lifetimes and 
the oscillation frequency in $B^0$-$\bar B^0$ mixing, which is characterized
by the mass-difference between the two mass eigenstates of the 
$B^0$-$\bar B^0$ system ($\Delta m_d$).

In this article, we report on the results of proper-time analyses of 
$B$ decays at Belle:
measurements of $\sin2\phi_1$\cite{CPV_BELLE}, the lifetime of $B$-mesons,
and $\Delta m_d$.

\section{ KEKB Accelerator and Belle }

 KEKB~\cite{KEKB} is an asymmetric $e^+e^-$ collider of 3~km circumference,
which consists of 8~GeV $e^-$ and 3.5~GeV $e^+$ storage rings and an injection
linear accelerator for them. 
It has one interaction point where the $e^+$ and
$e^-$ beams collide with a finite crossing angle of 22~mrad. 
The collider has been operated with peak beam currents of 1230~mA($e^+$) and 
780~mA($e^-$), giving a peak luminosity of 5.5$\times 10^{33}$/cm$^2$/sec
(as of the end of 2001, the target is 10$^{34}$/cm$^2$/sec).
Due to the energy asymmetry, 
the $\Upsilon(4S)$ and its daughter $B$-pair 
are produced with $\beta\gamma =$0.425 along the electron beam direction
(z direction) in the laboratory frame. 
The average distance between the two decay vertices of $B$ mesons is 
approximately 200~$\mu$m.
A total integrated luminosity
of 47~fb$^{-1}$ was accumulated during the period between October
1999 and the end of 2001, of which 4.0~fb$^{-1}$ was taken off-resonance
and the rest was taken on the $\Upsilon(4S)$.
 In the analyses reported here, we use a 29.1 fb$^{-1}$ data sample,
which contains 31.3 million $B \bar B$ pairs,
recorded at $\Upsilon (4S)$ between January 2000 and July 2001.

 Belle is an international collaboration consisting of $\sim$300 
physicists from $\sim$50 institutes in 14 countries (U.S.A., Russia, Asia, 
Australia, and Europe).
The Belle detector~\cite{Belle} is a general purpose large solid 
angle magnetic spectrometer surrounding the interaction point. 

Charged particle tracking is done
by a silicon vertex detector (SVD) and a central drift chamber (CDC).
The SVD consists of three layers of double-sided silicon strip 
detectors (DSSD) at radii of 3.0, 4.5 and 6.0~cm.
The CDC is a small-cell cylindrical drift 
chamber consisting of 50 layers of anode wires (18 stereo wire layers),
covering $17^\circ < \theta_{lab} < 150^\circ$.
The CDC is operated with a $He$(50\%)$+C_2H_6$(50\%) mixture.
The CDC also provides measurements of the energy loss 
with a resolution of $\sigma(dE/dx)$=6.9\%. 

Particle identification
is done by three detectors: $dE/dx$
measurements in the CDC, time-of-flight counters (TOF)  
and aerogel Cherenkov counters (ACC).  
The TOF system consists of 128 plastic scintillators. 
The time resolution is 95~psec ($rms$).
The ACC consists of 1188 aerogel blocks with refractive indices of between
1.01 and 1.03, depending on the polar angle. 
By combining information from these detectors, 
the efficiency
for $K^\pm$ identification is about 90\% and the $\pi$ fake rate is 
6\% with the requirement $P(K/\pi)>0.6$.

The electromagnetic calorimeter (ECL) consists of
8736 CsI(Tl) crystal blocks, 16.1 radiation length 
thick, and covering the same angular region as CDC.
%
Electron identification in Belle is based on
a combination of $dE/dx$ measurements
in the CDC, the response of the ACC, and the position, shape
and total energy (i.e. $E/p$) of its associated CsI shower.  

The outermost detector,
for the measurement of $\mu^\pm$ and $K_L$ (KLM), 
consists of 14 layers
of iron (4.7~cm thick) absorbers alternating with resistive plate counters 
(RPC). 

\section{Measurement of $\sin 2\phi_1$}

The Standard Model (SM) predicts
a $CP$ violating asymmetry in the time-dependent
rates for the initial $B^0$ and $\overline{B}{}^0$
decays to a common $CP$ eigenstate, $f_{CP}$.  
In the case where $f_{CP}=(c\overline{c})K^0$, the asymmetry
is given by
\begin{eqnarray}
A(\Delta t) &\equiv& 
  \frac{R(\overline{B}^0\rightarrow f_{CP})-R(B^0\rightarrow f_{CP})}
       {R(\overline{B}^0\rightarrow f_{CP})+R(B^0\rightarrow f_{CP})}
 \nonumber \\
 &=& -\xi_f\sin 2\phi_1\sin(\Delta m_d \Delta t),
\end{eqnarray}
where 
$\xi_f$ is the $CP$-eigenvalue of $f_{CP}$,
$\Delta t$ is proper-time difference of two $B$ decays,
and $\phi_1$  is one of the three internal
angles of the Unitarity Triangle, defined as
$\phi_1\equiv 
\pi-\arg\left(\frac{-V^*_{tb}V_{td}}{-V^*_{cb}V_{cd}}\right)
~(\equiv \beta)$~\cite{Sanda}.

%
The measurement requires the reconstruction of $B^0\to f_{CP}$
decays, the determination of the $b$-flavor of the accompanying 
$B$-decay ($f_{\rm tag}$), 
a measurement of $\Delta t$,
and a fit of the expected $\Delta t$ distribution to 
the measured distribution using a maximum-likelihood method.

We reconstruct $B^0$ decays to the following ${CP}$ eigenstates~\cite{CC}:
$J/\psi K_S$, $\psi(2S)K_S$, $\chi_{c1}K_S$, $\eta_c K_S$ for $\xi_f=-1$  and
$J/\psi K_L$ for $\xi_f=+1$.
The $J/\psi$ and  $\psi(2S)$ mesons are reconstructed via their decays to
$\ell^+\ell^-$ $(\ell=\mu,e)$.
The $\psi(2S)$ is also reconstructed via its  $J/\psi\pi^+\pi^-$ decay,
the $\chi_{c1}$ via its $J/\psi\gamma$ decay, and
the $\eta_c$ via its $K^+K^-\pi^0$
and $K_S(\pi^+\pi^-)K^-\pi^+$ decays.    
We also use $B^0\to J/\psi K^{*0} (K_S\pi^0)$ decays,
where the final state is a mixture of even
and odd $CP$. 
We find that the final state is primarily $\xi_f=+1$;
the $\xi_f = -1$ fraction is $0.19 \pm 0.04({\rm stat})\pm 
0.04({\rm syst})$~~\cite{Itoh}.

Except for $B^0\to J/\psi K_L$,  we identify $B$
decays using
the energy difference $\Delta E\equiv E_B^{\rm cms} - E_{\rm beam}^{\rm cms}$
and the beam-energy constrained
mass $M_{\rm bc}\equiv\sqrt{(E_{\rm beam}^{\rm cms})^2-(p_B^{\rm cms})^2}$,
where $E_{\rm beam}^{\rm cms}$ is the cms beam energy,
and $E_B^{\rm cms}$ and $p_B^{\rm cms}$ are the cms energy and momentum
of the $B$ candidate.
%
The $B$ meson signal region is defined as
$5.270<M_{\rm bc}<5.290~{\rm GeV}/c^2$ and 
a mode-dependent window around $\Delta E$ = 0.
Table~\ref{tab:tally} lists
the numbers of  observed candidates ($N_{\rm ev}$) and
the background ($N_{\rm bkgd}$) determined by
extrapolating the rate in the
non-signal  $\Delta E$ {\em vs.} $M_{\rm bc}$ region
into the signal region.

\begin{table}[!htbp] 
\caption{ Numbers of  observed
events ($N_{\rm ev}$) and the estimated 
background ($N_{\rm bkgd}$)
in the signal region for each $f_{CP}$ mode.}
\label{tab:tally}
\vspace{0.2cm}
\begin{center}
\footnotesize
\begin{tabular}{lrr}
\hline
Mode & $N_{\rm ev}$ & $N_{\rm bkgd}$\\
\hline
$J/\psi(\ell^+\ell^-) K_S(\pi^+\pi^-)$ & 457 & 11.9\\
$J/\psi(\ell^+\ell^-) K_S(\pi^0\pi^0)$  & 76 & 9.4\\
$\psi(2S)(\ell^+\ell^-)K_S(\pi^+\pi^-)$  & 39 & 1.2\\
$\psi(2S)(J/\psi\pi^+\pi^-)K_S(\pi^+\pi^-)$ & 46 & 2.1\\
$\chi_{c1}(J/\psi\gamma) K_S(\pi^+\pi^-)$ & 24 & 2.4\\
$\eta_c(K^+K^-\pi^0)K_S(\pi^+\pi^-)$ & 23 & 11.3\\
$\eta_c(K_S K^-\pi^+)K_S(\pi^+\pi^-)$ &41 & 13.6\\
$J/\psi K^{*0}(K_S\pi^0)$& 41 & 6.7\\
\hline
Sub-total & 747 & 58.6	\\
\hline
$J/\psi(\ell^+\ell^-) K_L$ & 569 & 223 \\
\hline
\end{tabular}
\end{center}
\end{table}

Candidate $B^0\to J/\psi K_L$  decays are selected by requiring
ECL and/or KLM hit patterns that are consistent with the 
presence of
a shower induced by a neutral hadron.
The quantity $p_B^{\rm cms}$ is
calculated with the $B^0 \to J/\psi K_L$ two-body decay hypothesis.
The $p_B^{\rm cms}$ distribution is fitted to the sum of signal
and background distributions.
There are 569 entries in the
$0.2\leq p_B^{\rm cms}\leq 0.45 (0.40)~~{\rm GeV}/c$
signal region with KLM (ECL) clusters. 

Tracks 
that are not associated with a reconstructed
$CP$ eigenstate decay are used for flavor tagging.
Initially,
the $b$-flavor determination is performed at the track level:
high momentum leptons
from  $b\to$ $c\ell^-\overline{\nu}$,
lower momentum leptons from  $c\to$ $s\ell^+\nu$,
charged kaons and $\Lambda$ baryons from $b\to$ $c\to$ $s$,
high momentum pions that originate from 
decays of $B^0\to$ $D^{(*)-}(n\pi)^+$,
and
slow pions from $D^{*+}\to D^0\pi^+$.
We use the MC to determine a category-dependent variable
that 
ranges from 
$-1$ for a reliably identified $\overline{B}{}^0$
to $+1$ for a reliably identified $B^0$, and  depends
on the tagging particle's  charge,
cms momentum, polar angle and
particle-identification probability, as well as other kinematic and
event shape quantities. 
The results from the separate
track categories are then combined to 
take into account correlations
in the case of multiple track-level tags.
This stage determines two event-level parameters, $q$ and $r$:
$q$ takes values $q = +1(-1)$  when the tag-side $B$ meson
is more likely to be a $B^0$
($\overline{B}{}^0$);
$r$ is an event-by-event flavor-tagging dilution factor, which
ranges from $r=0$ for no flavor discrimination to $r=1$ for
unambiguous flavor assignment.
It is used only to sort data into six intervals of $r$, according to 
the flavor purity.

The wrong-tag probabilities, $w_l\ (l=1,6)$,
are determined directly from the data for the six  $r$ intervals
using fully reconstructed, self-tagged
decays, as described in Sec.~5.1.
The total effective tagging efficiency is
$\sum_l f_l(1-2w_l)^2 = 
0.270\pm 0.008\rm{(stat)}^{+0.006}_{-0.009}\rm{(syst)}$,
where $f_l$ is the fraction of the events in each $r$ interval.

The vertex positions for the $f_{CP}$ and $f_{\rm tag}$ decays are
reconstructed using tracks
that have at least one
three-dimensional coordinate determined from 
associated $r$-$\phi$ and $z$
hits in the same SVD layer
along with one or more additional $z$ hits in the other layers.
Each vertex position is required to be
consistent with the interaction point profile smeared in the
$r$-$\phi$ plane by the $B$ meson decay length.
The $f_{CP}$ vertex is determined using
lepton tracks  from
$J/\psi$ or $\psi(2S)$ decays, or prompt tracks from $\eta_c$ decays.
The $f_{\rm tag}$ vertex
is determined from well reconstructed tracks not assigned to $f_{CP}$.
Tracks that form a $K_S$ are not used.

We determine  $\sin 2\phi_1$  by performing an
unbinned maximum-likelihood fit of a  
probability density function (pdf) to the observed $\Delta t$ distributions.
The pdf is a convolution of the theoretical distribution with
a $\Delta t$ resolution function, including the effect the of background
and wrong-tag probabilities.
The result is
 
$  \sin 2\phi_1 = 
     0.99 \pm 0.14 ({\rm stat}) \pm 0.06({\rm syst}). $

\begin{figure}[htp]
\begin{center}
\includegraphics[width=10cm]{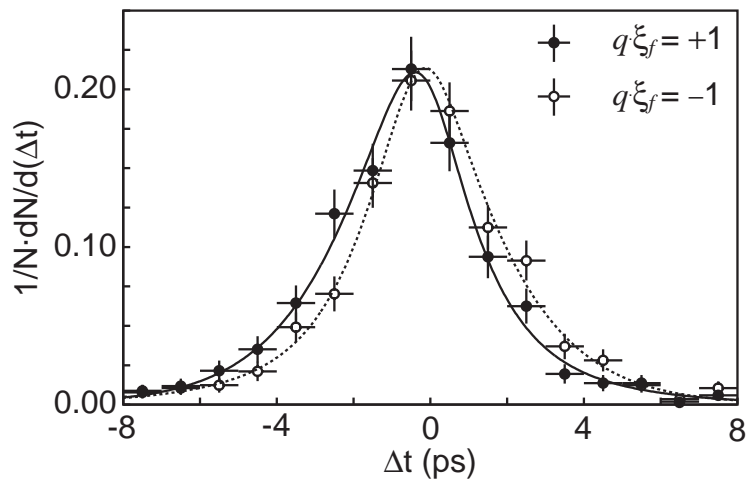}
\end{center}
\vskip -0.4cm  
\caption{
$\Delta t$ distributions 
for the events with $q\xi_f = +1$ (solid
points) and $q\xi_f = -1$ (open points). The 
results of the global fit (with  $\sin 2\phi_1 = 0.99$)
are shown as solid and dashed curves, respectively.
}
\label{fig:dNdt}
\end{figure}

\begin{figure}[!htpb] 
\begin{center}
\includegraphics[width=8cm,height=12.5cm]{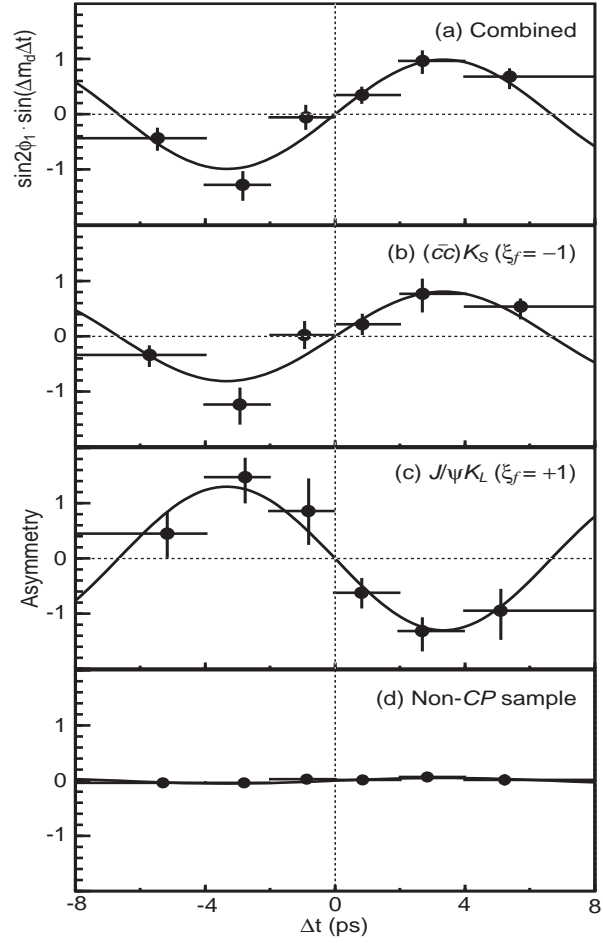}
\end{center}
\vskip -0.4cm 
\caption{(a) The asymmetry obtained
from separate fits to each $\Delta t$ bin for 
the full data sample; the curve is the result of 
the global fit. The
corresponding plots for the (b) $(c\bar{c})K_S$ ($\xi_f=-1$), (c) 
$J/\psi K_L$ ($\xi_f = +1$), and (d) $B^0$ control samples
are also shown.  The curves
are the results of the fit applied separately to the
individual data samples.
}
\label{fig:asym}
\end{figure}

\noindent
The systematic error is dominated by the uncertainties in vertex 
reconstruction and resolution.
Figure~\ref{fig:dNdt} shows the  
observed $\Delta t$ distributions
for the $q\xi_f =+1$ 
(solid points) and 
$q\xi_f = -1$ (open points) event samples.
In Fig.~\ref{fig:asym}(a) we show the 
asymmetries for the combined data sample 
that are obtained by applying the fit to the events in each
$\Delta t$ bin separately.  The smooth curve is the result
of the global unbinned fit.
Figures~\ref{fig:asym}(b) and (c) show the 
corresponding asymmetries
for the $ (c\bar{c})K_S$  ($\xi_f=-1$)  
and the $J/\psi K_L$ ($\xi_f=+1$) modes separately. 
The fits give $\sin 2\phi_1$ values to be
$0.84 \pm 0.17$(stat) and $1.31 \pm 0.23 $(stat), respectively,
which are opposite, as expected.
Fitting
to the non-$CP$ eigenstate self-tagged modes 
$B^0\to D^{(*)-}\pi^+$, $D^{*-}\rho^+$, $J/\psi 
K^{*0}(K^+\pi^-)$ and $D^{*-}\ell^+\nu$,
yields $0.05 \pm 0.04$.
As shown in Fig.~~\ref{fig:asym}(d),
 no asymmetry is seen, as expected.

\section{ Lifetime Measurements }

 For lifetime measurements, $\bar B^0$ and $B^-$ mesons are fully
reconstructed in the following hadronic decay modes:
$\bar B^0 \to D^+\pi^-$, $D^{*+}\pi^-$, $D^{*+}\rho^-$, $J/\psi K_S$,
$J/\psi K^{*0}(K^-\pi^+)$, and $B^- \to D^0\pi^-$, $J/\psi K^-$.
The reconstruction of $J/\psi X$ is made in the way quite similar to 
that used for the $CP$ analysis described above. 
Neutral and charged $D$ candidates are reconstructed
in the following channels:
$D^0 \to K^-\pi^+$, $K^-\pi^+\pi^0$, $K^-\pi^+\pi^+\pi^-$, 
and $D^+ \to K^-\pi^+\pi^+$.
$D^{*+} \to D^0\pi^+$ candidates are formed by combining
a $D^0$ candidate with a positively charged soft pion.
%
To reduce the continuum background, a selection based on the ratio of
the second to zeroth Fox-Wolfram moments\cite{FW} and the angle
between the thrust axes of the reconstructed and associated $B$ mesons
is applied mode by mode.
The vertices for fully reconstructed and associated $B$'s are reconstructed
using the same algorithm that is used in the $CP$ analysis.
We find 7863 $\bar B^0$ and 12047 $B^-$ events within the 
$\Delta E$-$M_{\rm bc}$
signal regions after the vertexing and selections are applied.

We extract the $\bar B^0$ and $B^-$ lifetimes by performing an unbinned
maximum-likelihood fit to the $\Delta t$ distributions of 
$\bar B^0$ and $B^-$ simultaneously.
The pdf is similar to the one used in the $CP$ analysis.
However, we have developed a more elaborate resolution function 
for lifetime measurements. 
It is constructed as the convolution of four different contributions:
the detector resolutions for the fully reconstructed $B$ and 
associated $B$,
the additional smearing from the associated $B$ due to the inclusion of 
secondary tracks
and the kinematic approximation that the $B$ mesons are at rest
in the cms.  We allow for outliers with a large-width Gaussian to
take into account any imperfection of the resolution function.
The fit includes 10 free parameters (7 for resolution function and 3
for outliers) besides $\tau_{\bar B^0}$ and $\tau_{B^-}$.
We obtain\cite{Belle_life}:

$  \tau_{\bar B^0} = 1.554 \pm 0.030({\rm stat}) \pm 0.019({\rm syst}) 
  ~{\rm ps}, $ 

$  \tau_{B^-} = 1.695 \pm 0.026({\rm stat}) \pm 0.015({\rm syst})
  ~{\rm ps}, $ 

$  \tau_{B^-}/\tau_{\bar B^0} = 
                  1.091 \pm 0.023({\rm stat}) \pm 0.014({\rm syst}).  $

\begin{figure}[!htb]
  \center
  \resizebox{0.7\textwidth}{!}{\includegraphics{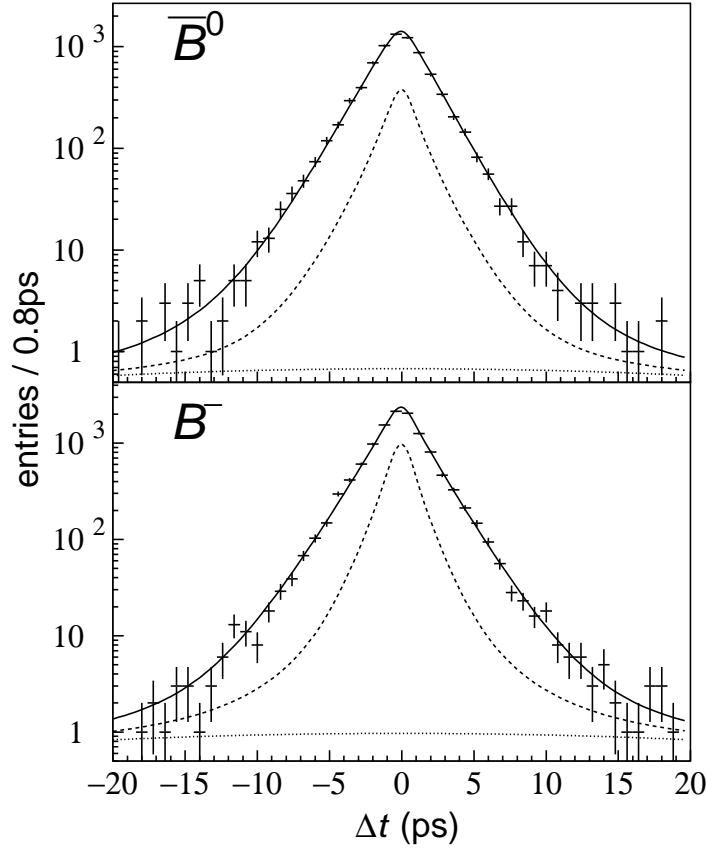}}
  \vskip -0.4cm  
  \caption{ $\Delta t$ distributions of neutral (top) and charged (bottom)
    $B$ meson pairs with fitted curves.  The dashed lines represent
    the sum of the background and the outlier component, and the dotted lines
    represent the outlier component.}
  \label{fig:fit_result}
\end{figure}

\noindent
Figure \ref{fig:fit_result} shows the distributions of $\Delta t$ for
$\bar B^0$ and $B^-$ events in the signal region with the fitted curves
superimposed.
The dominant sources of the systematic errors are the uncertainties in 
the background $\Delta t$ shape and the modeling of resolution function.

\section{ $\Delta m_d$ Measurements }

$\Delta m_d $ can be measured from the oscillation pattern of the
$\Delta t$ distributions of
the mixed (same flavor, SF) and unmixed (opposite flavor, OF) $B$ pairs:
\vspace{-0.1cm}
\begin{eqnarray}
N_{\rm SF} &\propto& e^{- |\Delta t|/\tau_{B^0}}
  \left[1 -\cos(\triangle m_d \Delta t) \right], \\ 
N_{\rm OF} &\propto& e^{- |\Delta t|/\tau_{B^0}}
  \left[1 +\cos(\triangle m_d \Delta t) \right].
\end{eqnarray}
\vspace{-0.1cm}
We published a result using inclusive dileptons based on an early
data sample\cite{DLmix}.
Here, we report on the results using three other different samples:
fully reconstructed hadronic modes, a fully reconstructed semi-leptonic
mode, and a partially reconstructed $D^{*\pm}\pi^\mp$ mode.

\subsection{ Fully Reconstructed Modes }

We use fully reconstructed, self-tagged decay modes in two categories:
hadronic decays ($D^{(*)-}\pi^+$, $D^{*-}\rho^+$) and
semi-leptonic decay ($B^0\to D^{*-}\ell^+\nu$).
For hadronic modes, the same event samples are used as lifetime measurements.
For $B^0\to D^{*-}\ell^+\nu$, we use the same $D^{*-}$ decay chains as
for the hadronic modes.
The $D^{*-}$ candidates are combined with $\mu^+$ or $e^+$
candidates having charge opposite to that of the $D^{*-}$ candidate.
We exploit the 
massless character of the neutrino
using 
the missing-mass squared in the cms defined by
$M_{\rm miss}^2 = MM - C \cos\theta_{B,D^*\ell}$ where
$MM = (E_B-E_{D^*\ell})^2-|\vec{p}_B|^2-|\vec{p}_{D^*\ell}|^2$,
$C = 2|\vec{p}_B|\,|\vec{p}_{D^*\ell}|$.
$E_B$ and $\vec{p}_B$ ($E_{D^*\ell}$ and  $\vec{p}_{D^*\ell}$)
are cms energy and momentum of 
$B^0$  
($D^*$ and lepton system).
$\theta_{B,D^*\ell}$ is an angle between vectors $\vec{p}_B$ and
$\vec{p}_{D^*\ell}$.  
It gives $\cos\theta_{B,D^*\ell} = - MM/C$ by setting $M_{\rm miss}^2 = 0$.
We select $D^{*-}\ell^+\nu$ candidates by
requiring  $|\cos\theta_{B,D^*\ell}| < 1.1$.
The $b$-flavor of the accompanying $B$ meson
is assigned according to the flavor-tagging algorithm described above.
The values of
$w_l$ are obtained from the amplitudes of the
time-dependent $B^0\overline{B}{}^0$ mixing oscillations:
$(N_{\rm OF} - N_{\rm SF})/(N_{\rm OF}+N_{\rm SF})
=(1-2w_l )\cos (\Delta m_d \Delta t)$.
We perform a maximum-likelihood fit to SF and OF $\Delta t$ distributions
simultaneously with $\Delta m_d$ and $w_l$'s as free parameters.
($\Delta m_d$ is fixed at the world average value~\cite{PDG} when we
determine $w_l$'s.)
We obtain 
$\Delta m_d = 0.521 \pm 0.017({\rm stat})~^{+0.011}_{-0.014}({\rm syst})$
~ps$^{-1}$ (prelim.) for hadronic modes and 
$\Delta m_d = 0.489 \pm 0.012({\rm stat})~^{+0.011}_{-0.014}({\rm syst})$
~ps$^{-1}$ (prelim.)  for semi-leptonic mode.
Dominant sources of systematic errors for hadronic modes are 
uncertainties in the resolution
function and the background components, including mixing. 
Those for the semi-leptonic mode are uncertainties in 
the effects of the backgrounds.

\subsection{ Partially Reconstructed $D^{*\pm}\pi^\mp$ }

In this method, $B^0 \rightarrow D^{*\pm}\pi^\mp$ decays are reconstructed
using only the fast pion ($\pi_f$) directly from the $B^0$ and
the soft pion ($\pi_s$) from $D^{*+} \rightarrow D^0 \pi^+$. 
Here, we use a method based on reconstructing the $D^0$ missing
mass.
We select a $\pi_f$ with cms momentum 2.05 $< p^*_{\pi_f} <$ 2.45 GeV/c
and an oppositely charged $\pi_s$ with a cms momentum of less than 0.45 GeV/c.
The missing mass is then calculated assuming the $B$ meson is 
at rest in the cms.
We select events with missing mass larger than 1.85 GeV/c$^2$
as $D^{*\pm}\pi^\mp$ candidates.
We require $R_2$ to be less than 0.5 to reduce the continuum background.
 In order to identify the flavor of the other $B$, we require an additional
high momentum lepton with a cms momentum larger than 1.1 GeV/c.
To reduce the background from incorrect tags (i.e. $\pi_f$ and leptons
from the same $B$), we further require 
$\cos\theta^*_{\ell\pi_f} > -0.8$ where $\theta^*_{\ell\pi_f}$ is the
angle between a $\pi_f$ and a lepton in the cms.
We select 3686 OF (1213 SF) candidates with 73\% (62\%) purity
after the missing mass cut and lepton-tag.
The $z$-vertex of leptons is determined from the
intersection of the lepton tracks with the profile of
$B^0_d$ decay vertices.  
  An unbinned maximum likelihood fit is performed to the OF and SF
$\Delta z$ distributions to extract $\Delta m_d$.
We obtain 
$\Delta m_d = 0.505 \pm 0.017({\rm stat}) \pm 0.020({\rm syst})$~ps$^{-1}$
(prelim.).
The dominant sources of systematic errors are uncertainties in the background
fraction and the resolution function.

\begin{figure}[phtb!]  
\begin{center}
 \resizebox{0.8\textwidth}{!}{\includegraphics{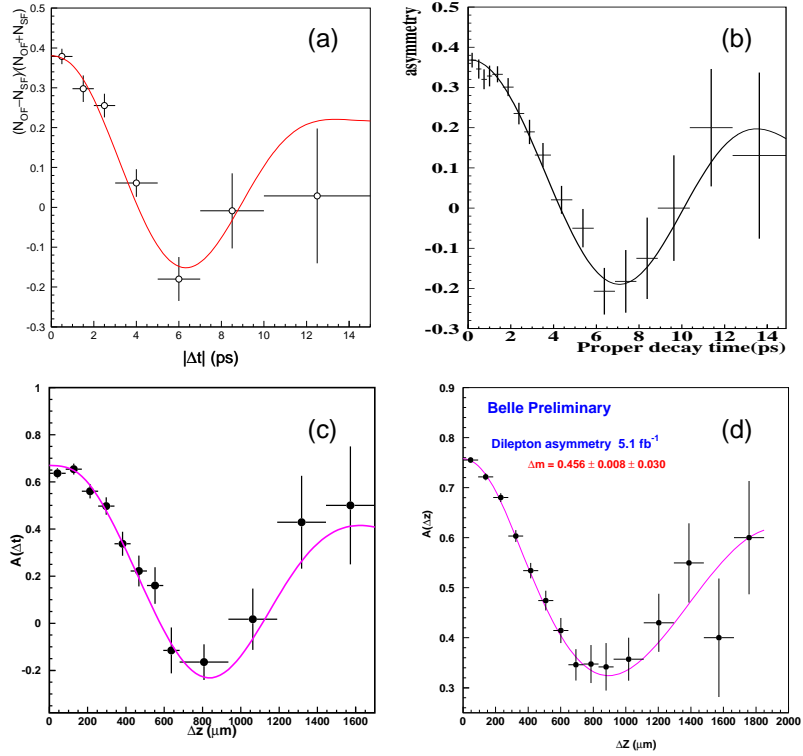}}
\vskip -0.4cm  
\caption{Asymmetry as a function of the proper decay time difference  
         for (a) hadronic modes, (b) semi-leptonic mode, 
         (c) partially reconstructed $B^0 \rightarrow D^*\pi$,
         and (d) dileptons.   
         Fit curves are also shown.}
\label {mix_asym}
\end{center}
\end{figure}

Figure~\ref{mix_asym} shows the OF-SF asymmetries as a function of
$\Delta z$ for each method mentioned above
together with the fit curves.

\section{ Conclusion }

Belle (along with BaBar) presented the first significant observation of
the $CP$ violation in 2001 summer.  The measured value is

$ \sin2\phi_1 = 0.99 \pm 0.14({\rm stst}) \pm 0.06({\rm syst}) $

\noindent
with more than 6$\sigma$ significance.
We have also measured the lifetimes of $\bar B^0$ and $B^-$ and $\Delta m_d$.
The results are consistent with those of other 
experiments~\cite{LIFE}\cite{MIX}. 
We find $\tau_{B^-}/\tau_{\bar B^0}$ is different
from 1 with more than 3$\sigma$ significance.
By combining four different methods, we obtain 
$\Delta m_d = 0.490 \pm 0.010 $~ps$^{-1}$ (prelim.).
These are the most precise measurements so far. 

\subsection*{ Acknowledgments }
We would like to acknowledge the KEKB accelerator group for 
excellent operation of the collider.
I would also like to thank the conference organizers for their
efforts.


\begin{thebibliography}{9}
\bibitem{CC}
 Throughout this article, 
 the inclusion of the charge conjugate mode is implied.
\end{thebibliography}
\end{document}